# Some Theoretical Results Regarding the Polygonal Distribution

Hien D. Nguyen* and Geoffrey J. McLachlan

January 17, 2017

### Abstract

The polygonal distributions are a class of distributions that can be defined via the mixture of triangular distributions over the unit interval. The class includes the uniform and trapezoidal distributions, and is an alternative to the beta distribution. We demonstrate that the polygonal densities are dense in the class of continuous and concave densities with bounded second derivatives. Pointwise consistency and Hellinger consistency results for the maximum likelihood (ML) estimator are obtained. A useful model selection theorem is stated as well as results for a related distribution that is obtained via the pointwise square of polygonal density functions.

## 1 Introduction

Piecewise approximation is among the core calculus techniques for the analysis of complex functions (cf. Larson & Bengzon (2013, Ch. 1)). In particular, piecewise-linear functions have been known to provide useful approximations in the analysis of curves and integrals; see for example the use of linear splines for functional approximation and the trapezoid method for quadrature in Sections 9.3 and 12.1 of Khuri (2003), respectively.

---

*HDN is at the Department of Mathematics and Statistics, La Trobe University, Melbourne. GJM is at the School of Mathematics and Physics, University of Queensland, St. Lucia. (*Corresponding author email: hien1988@gmail.com)

In statistics, there have been numerous uses of piecewise-linear functions as models for probability distributions and densities. The simplest of these distributions is the triangular distribution, which is often used as a teaching tool for understanding the fundamentals of distribution theory (cf. Doanne (2004) and Price & Zhang (2007)). Furthermore, triangular distributions have been used in numerous applications, including task completion time modeling in Program Evaluation and Review Techniques (PERT), as well as for the modeling of securities prices on the New York Stock Exchange, and haul times modeling of civil engineering data; see Kotz & Van Dorp (2004, Ch. 1) for details.

Beyond the simple triangular distribution, numerous piecewise-linear models have been proposed for the modeling of random phenomena. These include the Grenander estimator for monotone densities (Grenander, 1956), triangular kernel density estimators (Silverman, 1986, Ch.3), the disjointed piecewise-density estimators of Beirlant et al. (1999), the trapezoidal distributions of van Dorp & Kotz (2003), the general segmented distributions of Vander Wielen & Vander Wielen (2015), and the mixture of triangular distributions of Ho et al. (2016). Differing from the mixture formation of Ho et al. (2016), Karlis & Xekalaki (2001) considered the polygonal densities that are defined via the following characterization.

Let $X \in [0,1]$ be a random variable and let $Z \in [g]$ ($[g] = \{1, 2, ..., g\}$). Suppose that $\mathbb{P}(Z = i) = \pi_i \geq 0$ for $i \in [g]$ ($\sum_{i=1}^{g} \pi_i = 1$), and that the following conditional characterization holds for $i \in [g]$:

$$f(x|Z=i) = f(x;\theta_i) = \begin{cases} 2x/\theta_i, & \text{if } x < \theta_i, \\ 2(1-x)/(1-\theta_i), & \text{if } x \geq \theta_i, \end{cases} \quad (1)$$

if $\theta_i \in (0,1)$; $f(x|Z=i) = 2(1-x)$ if $\theta_i = 0$; and $f(x|Z=i) = 2x$ if $\theta_i = 1$. We call the distribution that is specified by the marginal density,

$$f(x;\boldsymbol{\theta}) = \sum_{i=1}^{g} \pi_i f(x|Z=i) = \sum_{i=1}^{g} \pi_i f(x;\theta_i), \quad (2)$$

a polygonal distribution with $g$-components and parameter $\boldsymbol{\theta} \in \Theta_g$, where

$$\Theta_g = \left\{ \boldsymbol{\theta}^\top = (\pi_1,...,\pi_g,\theta_1,...,\theta_g) : \pi_i \geq 0, \sum_{i=1}^{g} \pi_i = 1, \theta_i \in [0,1], \text{ and } i \in [g] \right\}.$$

In Karlis & Xekalaki (2008), it is shown that the polygonal distribution includes as special cases the uniform distribution, as well as the trapezoidal distributions of van Dorp & Kotz (2003). The recent reports of Kacker & Lawrence (2007) and Goncalves & Amaral-Turkman (2008) have demonstrated the potential use of trapezoidal distributions in metrology and genomics, respectively. Along with the recent efficient algorithm for ML (maximum likelihood) estimation of polygonal distributions by Nguyen & McLachlan (2016), we believe that an exploration of some theoretical properties of the polygonal distribution is now warranted as it has the potential to be a widely-applicable model for data analysis.

In this article, we deduce firstly that the polygonal distribution can be consistently estimated via ML estimation. Secondly, we show that the polygonal distribution is dense within the class of concave functions, with respect to the supremum norm. Thirdly, we compute bounds on the Hellinger divergence (cf.

Gibbs & Su (2002)) bracketing entropy for the polygonal distribution. These entropy bounds allows us to do two things. One, we can produce an exponential inequality that bounds the convergence of the ML function in Hellinger divergence. And two, the entropy bounds allow us to deduce a penalized likelihood-based model selection criterion for choosing the number of components for an optimal polygonal distribution fit.

## 2 Consistency of the Maximum Likelihood Estimator

Let $\boldsymbol{X}_n = \{X_j\}_{j=1}^n$ be an IID (independent and identically distributed) random sample from a polygonal distribution with parameter $\boldsymbol{\theta}_0^\top = \left(\pi_1^0, ..., \pi_g^0, \theta_1^0, ..., \theta_g^0\right) \in \Theta^0$, where

$$\Theta_g^0 = \left\{\boldsymbol{\theta} \in \Theta_g : E\left[L\left(\boldsymbol{\theta}; \boldsymbol{X}_n\right)\right] = \sup_{\boldsymbol{v} \in \Theta_g} E\left[L\left(\boldsymbol{v}; \boldsymbol{X}_n\right)\right]\right\}.$$

Let the log-likelihood function for the sample be

$$L\left(\boldsymbol{\theta}; \boldsymbol{X}_n\right) = \sum_{i=1}^n \log f\left(X_i; \boldsymbol{\theta}\right) = \sum_{i=1}^n \log \sum_{i=1}^g \pi_i f\left(X_i; \theta_i\right),$$

define the ML estimator as any $\hat{\boldsymbol{\theta}}_n \in \Theta_n$, where

$$\Theta_g^n = \left\{\boldsymbol{\theta} \in \Theta_g : L\left(\boldsymbol{\theta}; \boldsymbol{X}_n\right) = \sup_{\boldsymbol{v} \in \Theta_g} L\left(\boldsymbol{v}; \boldsymbol{X}_n\right)\right\}.$$

In Nguyen & McLachlan (2016), an efficient MM algorithm (minorization–maximization; see Hunter & Lange (2004)) was derived, for the computation of $\hat{\boldsymbol{\theta}}_n$, when $\boldsymbol{X}_n$ is fixed at some realized value. We now investigate the consistency of the ML estimator. Write the Euclidean-vector norm as $\|\cdot\|_2$; the following result is adapted from van der Vaart (1998).

**Lemma 1** (van der Vaart, 1998, Thm. 5.13). *Assume that (i) $\log f(x; \boldsymbol{\theta})$ is upper-semicontinuous, and that (ii) $\sup_{\boldsymbol{\theta} \in \mathbb{T}} \log f(X; \boldsymbol{\theta})$ is measurable and $E \sup_{\boldsymbol{\theta} \in \mathbb{T}} \log f(X; \boldsymbol{\theta}) < \infty$ for every sufficiently small ball $\mathbb{T} \subset \Theta_g$. If $\tilde{\boldsymbol{\theta}}_n$ is an estimator, such that $L\left(\tilde{\boldsymbol{\theta}}_n; \boldsymbol{X}_n\right) \geq L(\boldsymbol{\theta}_0; \boldsymbol{X}_n) - o(1)$, for some $\boldsymbol{\theta}_0 \in \Theta^0$, then for every $\epsilon > 0$ and compact set $\mathbb{K} \subset \Theta_g$,*

$$\lim_{n \to \infty} P\left(\inf_{\boldsymbol{\theta} \in \Theta_g^0} \left\|\tilde{\boldsymbol{\theta}}_n - \boldsymbol{\theta}\right\|_2 \geq \epsilon \text{ and } \tilde{\boldsymbol{\theta}}_n \in \mathbb{K}\right) = 0.$$

We observe that $f(X; \boldsymbol{\theta})$ is continuous for every $\boldsymbol{\theta} \in \Theta_g$ since it is the convex combination of continuous functions, and thus assumption (i) is fulfilled as continuity implies semicontinuity. Further, since $f(x; \boldsymbol{\theta})$ is also continuous in $X$ and has bounded support, we have the fulfillment of assumption (ii) since $f$ is continuous in both $x$ and $\boldsymbol{\theta}$, we have $\sup_{x \in [0,1]} \sup_{\boldsymbol{\theta} \in \mathbb{T}} \log f(x; \boldsymbol{\theta}) < \infty$ for every ball $\mathbb{T} \in \Theta$, and thus $E \sup_{\boldsymbol{\theta} \in \mathbb{T}} \log f(X; \boldsymbol{\theta}) < \infty$ for every underlying polygonal distribution. Lastly, we can take $\tilde{\boldsymbol{\theta}}_n = \hat{\boldsymbol{\theta}}_n$ as it fulfills the condition of Lemma 1. We therefore have the following result.

**Theorem 1.** *The ML estimator $\hat{\boldsymbol{\theta}}_n \in \Theta_g^n$ is consistent in the sense that*

$$\lim_{n \to \infty} P\left(\inf_{\boldsymbol{\theta} \in \Theta_g^0} \left\|\hat{\boldsymbol{\theta}}_n - \boldsymbol{\theta}\right\|_2 \geq \epsilon \text{ and } \hat{\boldsymbol{\theta}}_n \in \mathbb{K}\right) = 0,$$

*for every $\epsilon > 0$ and $\mathbb{K} \subset \Theta_g$.*

*Remark* 1. Theorem 1 makes two allowances that are important for mixture models, such as the polygonal distribution. First, the ML estimator $\hat{\boldsymbol{\theta}}_n$ can be one of many possible global maximizers of the log-likelihood function. This is important as the lack of identifiability beyond a permutation of the component labels (cf. Titterington et al. (1985, Sec. 3.1)) guarantees that if global max-ima of the log-likelihood exists then there are at least $g!$ of them. Secondly, and similarly, since the lack of identifiability beyond a permutation also implies that for any given generative density function $f(x;\boldsymbol{\theta}_0)$, there are at least $g! - 1$ other models, which can be obtained by permuting the component labels of the parameter elements, that would generate the same random variable. Thus, the allowance for the generative model to arise from a set of possible parameter values, rather than a singular parameter value, provides a broader but more useful form of consistency in the context of mixture models. We note that this form of consistency is equivalent to the quotient-space consistency of Redner (1981) and Redner & Walker (1984), as well as the extremum estimator consistency of Amemiya (1985).

## 3 Approximations via Polygonal Distributions

We start by noting that every density of form (1) is a concave function in $x \in [0, 1]$, and thus we have the fact that (2) is also concave in $x$, by composition.

**Proposition 1.** *For any $\boldsymbol{\theta} \in \Theta_g$ and $g \in \mathbb{N}$, the polygonal distribution of form (2) is concave over the domain of $x \in [0, 1]$.*

*Proof.* Since each triangular component of form (1) is concave, and since (2) is a convex composition of triangular components, we obtain the desired result due to the preservation of concavity under positive summation (cf. Boyd & Vandenberghe (2004, Sec. 3.2)). □

We next consider regular piecewise-linear approximations of twice continuously-differentiable, positive, and concave functions $h$ (on $x \in [0, 1]$). That is, we consider approximations of the form

$$h(x) \approx l(x) = \left\{ h(t_i) + \frac{h(t_i) - h(t_{i+1})}{t_i - t_{i+1}} (x - t_i) \quad \text{if } t_i \leq x \leq t_{i+1}, \right. \tag{3}$$

for $i \in [g]$, where $t_i = (i-1)/g$. The following result is well known, although we follow the reporting of de Boor (2001, Ch. 3).

**Lemma 2** (de Boor, 2000, Ch. 3). *For any twice continuously-differentiable function $h$ over $x \in [0,1]$, the piecewise-linear approximation $l(x)$ (defined by (3)) satisfies the bound*

$$\sup_{x \in [0,1]} |h(x) - l(x)| \leq \frac{1}{8g^2} \sup_{x \in [0,1]} |h''(x)|. \tag{4}$$

Define a polygonal approximation as any function of form (2) over $x \in [0,1]$, where $\boldsymbol{\theta} \in \tilde{\Theta}_{g+1}$ and

$$\tilde{\Theta}_g = \left\{ \boldsymbol{\theta}^\top = (\pi_1, ..., \pi_{g+1}, \theta_1, ..., \theta_g) : \pi_i \geq 0, \theta_i \in [0,1], \text{ and } i \in [g] \right\}.$$

We seek to demonstrate that polygonal approximations can be used to construct a piecewise-linear approximation for any continuous, non-negative, and concave function $h$.

Start by setting $\theta_i = (i-1)/g$, for each $i \in [g+1]$. Note the implication that the polygonal approximation $f$ is discontinuous only at the nodes $\theta_i$. In Karlis & Xekalaki (2008), it is observed that

$$f(x) = 2(1-x)\left(\pi_1 + \sum_{i=2}^{k} \frac{\pi_i}{1-\theta_i}\right) + 2x\left(\sum_{i=k+1}^{g} \frac{\pi_i}{\theta_i} + \pi_{g+1}\right),$$

is linear in every interval $[\theta_k, \theta_{k+1}]$, for $k \in [g]$. For brevity, the summations are taken to be zero if the end points are incoherent.

Noting that $l(\theta_i) = h(\theta_i)$, for each $i \in [g+1]$, it suffices to match the values of $f(x)$ at each $\theta_i$, since the line segment between any two points in a Cartesian space is uniquely determined. Thus, we are required to solve the system of

equations

$$h(0) = 2\pi_1, \, h(1) = 2\pi_{g+1},$$

and

$$\begin{aligned} h\left(\frac{k-1}{g}\right) &= 2(g-k+1)\left[\frac{h(0)}{2g} + \sum_{i=2}^{k}\frac{\pi_i}{g-i+1}\right] \\ &\quad + 2(k-1)\left[\sum_{i=k+1}^{g}\frac{\pi_i}{i-1} + \frac{h(1)}{2g}\right], \end{aligned} \quad (5)$$

for each $k \notin \{1, g+1\}$. This is a linear system of full rank and thus is solvable. We must finally validate that the solution is non-negative under our hypothesis (i.e. $\pi_i^* \geq 0$ solves the system, for $i \in [g]$). Since $h(0) \geq 0$ and $h(1) \geq 0$, we have $\pi_1^* \geq 0$ and $\pi_{g+1}^* \geq 0$. For any $i \notin \{1, g+1\}$, we can check that

$$\pi_i^* = \frac{(g-i+1)(i-1)}{2g}\left[2h\left(\frac{i-1}{g}\right) - h\left(\frac{i}{g}\right) - h\left(\frac{i-2}{g}\right)\right]$$

solves (5). We obtain the desired outcome by noting that the definition of concavity implies $h([a+b]/2) \geq [h(a) + h(c)]/2$, for $0 \leq a < b \leq 1$. Via Lemma 2, we can now state our approximation theorem.

**Theorem 2.** *The class of of polygonal approximations is dense within the class of twice continuously-differentiable, non-negative, and concave functions with bounded second derivative, over the unit interval. That is, for any twice continuously-differentiable, non-negative, and concave function h with bounded second derivative, over the unit interval, there exists a $g \in \mathbb{N}$ and $f(x; \boldsymbol{\theta})$ (with $\boldsymbol{\theta} \in \tilde{\Theta}_{g+1}$), such that for every $\epsilon > 0$, $\sup_{x \in [0,1]} |h(x) - f(x; \boldsymbol{\theta})| < \epsilon$.*

*Proof.* For any $h$, we have a piecewise-linear approximation $l$, such that (4) holds. We have also shown that under the hypothesis, for any $l$ with $g$ segments, there exists a polygonal approximation with $g+1$ components that is equal to $l$. Thus, one can replace $l$ with $f$ in (4) and note that the second derivative is bounded in order to obtain the result. □

*Remark* 2. The obtained result is a finite approximation theorem. That is, for any level of accuracy $\epsilon > 0$, there exists a polygonal approximation with a finite number of components $g+1 < \infty$ that can provide the require degree of accuracy. This makes Theorem 2 comparable to denseness results for mixtures of shifted and scaled densities, such as DasGupta (2008, Thm. 33.2). One advantage that Theorem 2 has over DasGupta (2008, Thm. 33.2) and the likes is that we have been able to establish the component weights must be non-negative (i.e. $\pi_i \geq 0$, $i \in [g+1]$). In comparison, no such guarantees are made by DasGupta (2008, Thm. 33.2); see also the denseness results in Cheney & Light (2000).

We consider in passing a related family of distributions that can be obtained via the pointwise square of a polygonal approximation. That is define the family of density functions $\mathcal{S}$, where

$$\mathcal{S} = \left\{ s(x) = f^2(x; \boldsymbol{\theta}) : \boldsymbol{\theta} \in \tilde{\Theta}_{g+1}, \|f(x; \boldsymbol{\theta})\|_2^2 = 1, \text{ and } g \in \mathbb{N} \right\}$$

and $\|h(x)\|_2 = \left( \int_0^1 h^2(x) \, \mathrm{d}x \right)^{1/2}$ is the $\mathcal{L}_2$-norm over the unit interval. Define the KL divergence (Kullback-Leibler; Kullback & Leibler (1951)) from a density $f$ to $h$ (defined on $x \in [0,1]$) as $K(h\|f) = \int_0^1 h(x) \log \frac{f(x)}{h(x)} \mathrm{d}x$. The following specialization of Massart (2007, Prop. 7.20) allows for the use of Theorem 2 to derive an approximation theorem for densities in class $\mathcal{S}$. As the specialization is simple, we provide it without proof.

**Lemma 3** (Massart, 2007, Prop. 7.20)**.** *Let $\bar{\mathcal{T}}^{1/2}$ be a convex cone in $\mathcal{L}_\infty$ such that $1 \in \bar{\mathcal{T}}^{1/2}$ and let $\mathcal{T}^{1/2}$ be the set of elements from $\bar{\mathcal{T}}^{1/2}$ with $\mathcal{L}_2$-norm equal* to 1. If $\mathcal{T} = \{t^2 : t \in \mathcal{T}^{1/2}\}$, then for any density $h$

$$\min\left\{1, \inf_{t \in \mathcal{T}} K(h||t)\right\} \leq 12 \inf_{u \in \bar{\mathcal{T}}^{1/2}} \sup_{x \in [0,1]} \left|h^{1/2}(x) - u(x)\right|^2.$$

**Proposition 2.** *For any twice continuously-differentiable, non-negative, and concave density function $h$ with bounded second derivative (on $x \in [0,1]$), there exists a density $s \in \mathcal{S}$, such that for any $\epsilon \in (0,1)$, $\inf_{s \in \mathcal{S}} K(h||s) \leq \epsilon$.*

*Proof.* Define $\bar{\mathcal{S}}^{1/2}$ to be the set of polygonal approximations with potentially any $g \in \mathbb{N}$. Since the polygonal approximations are continuous on a bounded domain, they are elements of $\mathcal{L}_\infty$. Further, the sum and positive product of any two polygonal approximation is another polygonal approximation, thus the class forms a convex cone. The element 1 is in $\bar{\mathcal{S}}^{1/2}$ since the polygonal distributions includes the uniform distribution. Next, define $\mathcal{S}^{1/2}$ be the set of elements from $\bar{\mathcal{S}}^{1/2}$ with $\mathcal{L}_2$-norm equal to 1, and subsequently $\mathcal{S} = \{t^2 : t \in \mathcal{S}^{1/2}\}$. Since concavity is conserved under increasing and concave composition (cf. Boyd & Vandenberghe (2004, Eqn. 3.10)), the function $h^{1/2}$ is concave under the hypothesis. Thus, by Theorem 2, there exists a $u \in \bar{\mathcal{S}}^{1/2}$ such that

$$\inf_{u \in \bar{\mathcal{S}}^{1/2}} \sup_{x \in [0,1]} \left|h^{1/2}(x) - u(x)\right| \leq \delta$$

for any $\delta > 0$. We can then take any $\delta$ that makes $\epsilon = 12\delta^2 < 1$ to complete the proof. □

## 4 Hellinger Bracketing Results

We established the consistency of the ML estimator with respect to convergence in probability of the parameter estimate $\hat{\boldsymbol{\theta}}_n \in \Theta_g^n$ to some element within the set of maximizers of the expected log-likelihood $\Theta_g^0$. Since the log-likelihood is con-tinuous, continuous mapping can be used to obtain convergence in probability between the log-likelihood evaluated at the ML estimate and the log-likelihood evaluated at an element in $\Theta_g^0$, pointwise, for any $x \in [0, 1]$.

It is often more interesting to obtain a stronger mode of convergence. Convergence in KL divergence is one such mode, as is the related Hellinger divergence. For any two densities $f$ and $h$ (over $x \in [0, 1]$), we can define the squared-Hellinger divergence as $H^2(f, h) = \int_0^1 \left(f^{1/2} - h^{1/2}\right)^2 \mathrm{d}x$.

Define the $\mathcal{L}_2$ $\epsilon$-bracketing metric entropy of a class $\mathcal{T}$ as $\log N_B(\mathcal{T}, \mathcal{L}_2(\mu), \epsilon)$, where $\mu$ is the Lebesgue measure over the unit interval. The function is the $N_B(\mathcal{T}, \mathcal{L}_2(\mu), \epsilon)$ $\epsilon$-bracketing number and is equal to the smallest number of intervals of the form $[t, u]$ $(t, u \in \mathcal{L}_2)$, such that for any $v \in \mathcal{T}$, there exists an interval where $v \in [t, u]$, and every interval satisfies the bound $\|t - u\|_2 \leq \epsilon$, for $\epsilon > 0$ (cf. van der Vaart (1998, Sec. 19.2)). The following theorem of Gine & Nickl (2015) provides a bounding result for the ML estimator with respect to the Hellinger divergence.

**Lemma 4** (Gine and Nickl, 2015, Thm. 7.2.1). *Let $\boldsymbol{X}_\mathbb{N} = \{X_j\}_{j \in \mathbb{N}}$ be an infinite IID random sample with joint probability distribution $P_0^\mathbb{N}$ and marginal probability density $t_0 \in \mathcal{T}$. Let $\boldsymbol{X}_n = \{X_j\}_{j=1}^n$ be the first $n \in \mathbb{N}$ elements of $\boldsymbol{X}_\mathbb{N}$ and define $\hat{t}_n \in \mathcal{T}$ to be the ML estimator that satisfies the condition $\sup_{t \in \mathcal{T}} L(t; \boldsymbol{X}_n) = L(\hat{t}_n; \boldsymbol{X}_n)$, where $L(t; \boldsymbol{X}_n) = \sum_{j=1}^n \log t(X_j)$ is the log-likelihood function indexed by $t$. Let*

$$\bar{\mathcal{T}} = \left\{\bar{t} = \frac{t + t_0}{2} : t \in \mathcal{T}\right\} \text{ and } \bar{\mathcal{T}}^{1/2} = \left\{t^{1/2} : t \in \bar{\mathcal{T}}\right\}.$$

*If we take $J(\delta) \geq \max\left\{\delta, \int_0^\delta \sqrt{\log N_B(\bar{\mathcal{T}}^{1/2}, \mathcal{L}_2(\mu), \epsilon)} \, d\epsilon\right\}$ (for $\delta \in (0, 1]$) such that $J(\delta)/\delta^2$ is non-increasing, then there exists a fixed constant $C_1 > 0$ such* that for any $\delta_n^2$ that satisfies $\sqrt{n}\delta_n^2 \geq C_1 J(\delta_n)$, we have for all $\delta \geq \delta_n$,

$$P_0^{\mathbb{N}}\left(H\left(\hat{t}_n, t_0\right) \geq \delta\right) \leq C_1 \exp\left(-n\delta^2/C_1^2\right).$$

Using the notation of Lemma 4, we first define $\mathcal{T}$ to be the class of polygonal densities

$$\mathcal{T} = \mathcal{F}_g = \{f(x;\boldsymbol{\theta}) : \boldsymbol{\theta} \in \Theta_g\}$$

and suppose that we observe an IID sample generated from a distribution with marginal density $f_0 \in \mathcal{T}$. We are reminded that any $f \in \mathcal{T}$ is concave via Proposition 1. Now, consider that any element $\bar{f} \in \bar{\mathcal{T}}$ or $\bar{f}^{1/2} \in \bar{\mathcal{T}}^{1/2}$ must be concave, since affine combinations and positive concave compositions preserve concavity, respectively. The following adaptation of a result from Doss & Wellner (2016) provides a simple method for computing the $\epsilon$-bracketing metric entropy of the class $\bar{\mathcal{T}}^{1/2}$.

**Lemma 5** (Doss and Wellner, 2016, Prop. 4.1). *Let $\mathcal{C}([a,b], B)$ be the class of concave (or convex) functions over the interval $[a,b]$, such that for any $c \in \mathcal{C}([a,b], B)$, $|c(x)| \leq B$, for any $x \in [a,b]$. If $\epsilon > 0$ and $1 \leq r < \infty$, then there exists a constant $C_2 < \infty$ such that*

$$\log N_B\left(\mathcal{C}([a,b], B), \mathcal{L}_r(\mu), \epsilon\right) \leq C_2 \left[\frac{B(b-a)^{1/r}}{\epsilon}\right]^{1/2}.$$

Note that any $f \in \mathcal{T}$ is bounded by 2 and thus $\bar{f}^{1/2} \leq \sqrt{2}$. Set $B = \sqrt{2}$, $[a,b] = [0,1]$, and $r = 2$ and observe that the class of $\bar{\mathcal{T}}^{1/2}$ requires a smaller bracketing than does $\mathcal{C}\left([0,1], \sqrt{2}\right)$ as it is a proper subset. We thus have our desired bracketing entropy

$$\log N_B\left(\bar{\mathcal{T}}^{1/2}, \mathcal{L}_2(\mu), \epsilon\right) \leq \log N_B\left(\mathcal{C}\left([0,1], \sqrt{2}\right), \mathcal{L}_2(\mu), \epsilon\right) \leq 2^{1/4} C_2 \left(\frac{1}{\epsilon}\right)^{1/2}.$$

We can evaluate

$$\int_0^\delta \sqrt{\log N_B\left(\bar{\mathcal{T}}^{1/2}, \mathcal{L}_2(\mu), \epsilon\right)}\mathrm{d}\epsilon = \frac{2^{17/8}C_2^{1/2}}{3}\delta^{3/4}$$

and set $J(\delta) = \max\left\{\delta, \frac{2^{17/8}C_2^{1/2}}{3}\delta^{3/4}\right\}$. We can then verify that $J(\delta)/\delta^2$ is non-increasing since both of its components $1/\delta$ and $2^{17/8}C_2^{1/2}/\left(3\delta^{5/4}\right)$ are decreasing. We thus have the following result regarding the convergence in Hellinger divergence of the ML estimator.

**Proposition 3.** *Let $\boldsymbol{X}_\mathbb{N} = \{X_j\}_{j\in\mathbb{N}}$ be an infinite IID random sample with joint probability distribution $P_0^\mathbb{N}$ and marginal probability density $f_0 \in \mathcal{F}_g$. Let $\boldsymbol{X}_n = \{X_j\}_{j=1}^n$ be the first $n \in \mathbb{N}$ elements of $\boldsymbol{X}_\mathbb{N}$ If $\hat{f}_n = f\left(x;\hat{\boldsymbol{\theta}}_n\right) \in \mathcal{F}_g$ to be the ML estimator, then there exists two constants $C_1, C_2' > 0$ such that for any $\delta_n^2$ that satisfies*

$$\sqrt{n}\delta_n^2 \geq C_1 \max\left\{\delta_n, C_2'\delta_n^{3/4}\right\}, \tag{6}$$

*for all $n \in \mathbb{N}$, we have for all $\delta \geq \delta_n$,*

$$P_0^\mathbb{N}\left(H\left(\hat{f}_n, f_0\right) \geq \delta\right) \leq C_1 \exp\left(-n\delta^2/C_1^2\right). \tag{7}$$

*Remark* 3. Observe that when $\delta$ is fixed and $n$ is allowed to increase, (7) permits the usual exponential inequality argument for the diminishing probability of large deviation between $\hat{f}_n$ and $f_0$. Unfortunately, since we can only select $\delta \geq \delta_n$ for each $n$, where $\delta_n$ satisfies (6), we are not in a position to select arbitrarily small $\delta$ to guarantee small probabilities of Hellinger divergence, especially for small $n$. However, since the left hand side of (6) is increasing at a more rapid rate than the right, we are able to make stronger arguments regarding the closeness of the densities at larger values of $n$. Unfortunately, at which large value of $n$ is difficult to know as we must still contend with the two unknown constants $C_1$ and $C_2'$.

We now proceed to derive the entropy of any polygonal density function in $\mathcal{F}_g$, rather than the entropy of elements in $\bar{\mathcal{T}}^{1/2}$. Of course, we could simply reapply Lemma 5 to obtain such a value. However, for our intended application, we prefer a more nuanced approach that will allow us to obtain an entropy expression with a dependence on the number of components $g$. Define $\mathcal{F}_g^{1/2} = \{f^{1/2} : f \in \mathcal{F}_g\}$. The following result of Genovese & Wasserman (2000) provides a useful method for computing the entropy of finite mixture models; see also Maugis & Michel (2011).

**Lemma 6** (Genovese and Wasserman, 2000, Thm. 2). *Let*

$$\mathcal{P}_{g-1} = \left\{(\pi_1, ..., \pi_g) : \pi_i \geq 0 \text{ and } \sum_{i=1}^{g} = 1,\ i \in [g]\right\}$$

*be the probability simplex on $g$ points and let $\mathcal{K}$ be a class of component density functions, and let $\mathcal{P}_{g-1}^{1/2} = \{p^{1/2} : p \in \mathcal{P}\}$ and $\mathcal{K}^{1/2} = \{f^{1/2} : f \in \mathcal{K}\}$. If we define the mixture of $g$ densities from $\mathcal{K}$ to be*

$$\mathcal{M}_g = \left\{f(x) = \sum_{i=1}^{g} \pi_i f_i : f_i \in \mathcal{K}, (\pi_1, ..., \pi_g) \in \mathcal{P}_{g-1},\ \text{and}\ i \in [g]\right\}$$

*then*

$$N_B\left(\mathcal{M}_g^{1/2}, \mathcal{L}_2(\mu), \epsilon\right) \leq N_B\left(\mathcal{P}_{g-1}^{1/2}, \mathcal{L}_2(\mu), \frac{\epsilon}{3}\right) \left[N_B\left(\mathcal{K}^{1/2}, \mathcal{L}_2(\mu), \frac{\epsilon}{3}\right)\right]^g$$

*where $\mathcal{M}_g^{1/2} = \{f^{1/2} : f \in \mathcal{M}\}$ and*

$$N_B\left(\mathcal{P}_{g-1}^{1/2}, \mathcal{L}_2(\mu), \frac{\epsilon}{3}\right) \leq g(2\pi e)^{g/2} \left(\frac{3}{\epsilon}\right)^{g-1}.$$

Let $\mathcal{F} = \{f : f = f(x; \theta) \text{ has form (1) with } \theta \in [0, 1]\}$ be the class of trian-

gular density functions, and let $\mathcal{F}^{1/2} = \{f^{1/2} : f \in \mathcal{F}\}$. We apply Lemmas 5 and 6 to obtain the expression

$$\begin{aligned}
\log N_B \left( \mathcal{F}_g^{1/2}, \mathcal{L}_2(\mu), \epsilon \right) &\leq \log N_B \left( \mathcal{P}_{g-1}^{1/2}, \mathcal{L}_2(\mu), \frac{\epsilon}{3} \right) + g \log N_B \left( \mathcal{F}^{1/2}, \mathcal{L}_2(\mu), \frac{\epsilon}{3} \right) \\
&\leq \log(g) + \frac{g}{2} \log(2\pi e) + (g-1) \log\left(\frac{3}{\epsilon}\right) + g 2^{1/4} C_3 \left(\frac{3}{\epsilon}\right)^{1/2}
\end{aligned}$$

since the triangular density functions are bounded by 2. Using the fact that $\log x < \sqrt{x}$ for all $x > 0$ and letting $C_4(g) = \log(g) + g \log(2\pi e)/2$, we have

$$\log N_B \left( \mathcal{F}_g^{1/2}, \mathcal{L}_2(\mu), \epsilon \right) \leq g \left( 2^{1/4} C_3 + 1 \right) \left(\frac{3}{\epsilon}\right)^{1/2} + C_4(g). \tag{8}$$

Let $\mathcal{N} = \{\mathcal{F}_g : g \in [\gamma]\}$ be the set of families of mixtures of polygonal densities with up to $\gamma \in \mathbb{N}$ components. We shall now utilize (8) along with the following result from Massart (2007) in order to derive a penalized ML estimation method for selecting the optimal model class in $\mathcal{N}$. Our presentation follows the exposition of Maugis & Michel (2011).

**Lemma 7** (Massart, 2003, Thm. 7.11). *Let $\boldsymbol{X}_n = \{X_j\}_{j=1}^n$ be an IID random sample from some distribution with unknown density $t_0$. Let $\mathcal{N} = \{\mathcal{M}_g : g \in [\gamma]\}$ be some countable collection of models ($\gamma \in \mathbb{N}$) and let $\hat{t}_n^g \in \mathcal{M}_g$ be the ML estimator in the sense that $\sup_{t \in \mathcal{M}_g} L(t; \boldsymbol{X}_n) = L(\hat{t}_n^g; \boldsymbol{X}_n)$, where $L(t; \boldsymbol{X}_n) = \sum_{j=1}^n \log t(X_j)$ is the log-likelihood function indexed by $t$. Let $\rho_g$ for $g \in [\gamma]$ be some family of non-negative numbers satisfying*

$$\sum_{g=1}^\gamma \exp(-\rho_g) = \Sigma < \infty.$$

*Assume that for every $\mathcal{M}_g$, we have some non-decreasing function $J_g(\delta)$, such*

that $J_g(\delta)/\delta$ is non-increasing for $\delta \in (0, \infty)$ and

$$\int_0^\delta \sqrt{\log N_B\left(\mathcal{M}_g^{1/2}, \mathcal{L}_2(\mu), \epsilon\right)} \leq J_g(\delta).$$

Denote $\delta_g$ to be the unique positive solution of the equation $\sqrt{n}\delta^2 = J_g(\delta)$. If we let

$$crit(g) = L\left(\hat{t}_n^g; \boldsymbol{X}_n\right) + pen(g)$$

be a penalized ML criterion that we wish to minimize, then there exists some absolute constants $\kappa$ and $C_4$, such that whenever

$$pen(g) \geq \kappa\left(\delta_g^2 + \frac{\rho_g}{n}\right)$$

and some random $\hat{g}$ that minimizes crit over $\mathcal{N}$ exists, we have

$$E\left[H^2\left(t_0, t_n^{\hat{g}}\right)\right] \leq C_4\left[\inf_{g \in [\gamma]}\left[\inf_{t \in \mathcal{M}_g} K(t_0||t) + pen(g)\right] + \frac{\Sigma}{n}\right],$$

regardless of whether or not $t_0$ is in $\mathcal{N}$.

Referring back to the polygonal densities, we note that

$$\sqrt{\log N_B\left(\mathcal{M}_g^{1/2}, \mathcal{L}_2(\mu), \epsilon\right)} \leq \sqrt{\left[g\left(2^{1/4}C_3 + 1\right) - 1\right]\left(\frac{3}{\epsilon}\right)^{1/2} + \sqrt{C_4(g)}}$$

and thus

$$\int_0^\delta \sqrt{\log N_B\left(\mathcal{M}_g^{1/2}, \mathcal{L}_2(\mu), \epsilon\right)} d\epsilon \leq \frac{4}{3^{4/3}}\sqrt{g\left(2^{1/4}C_3 + 1\right)}\delta^{3/4} + \delta\sqrt{C_4(g)}.$$

Setting $J_g(\delta) = \frac{4}{3^{4/3}}\sqrt{g\left(2^{1/4}C_3 + 1\right)}\delta^{3/4} + \delta\sqrt{C_4(g)}$, we see that $J_g(\delta)$ is increasing and $J_g(\delta)/\delta \propto \delta^{-1/4}$ is decreasing. We also get the unique positive

solution
$$\delta_g = \left(\frac{C_5 g^{1/2}}{\left[n^{1/2} - C_4^{1/2}(g)\right]^3}\right)^{4/3}$$

to the equation $\sqrt{n}\delta^2 = J_g(\delta)$, where $C_5 = \frac{4}{3^{4/3}}\sqrt{2^{1/4}C_3 + 1}$. Next. we can select $\rho_g = g$ for each $g \in [\gamma]$ as it is representative of the complexity of each of the increasing number of components. Thus, for any finite $\gamma$, we have

$$\Sigma = \Sigma_\gamma = \frac{\exp(-1) - \exp(-\gamma - 1)}{1 - \exp(-1)}.$$

We can now apply Lemma 7 to obtain our model selection theorem for the polygonal distributions.

**Theorem 3.** *Let $\mathbf{X}_n = \{X_j\}_{j=1}^n$ be an IID random sample from some distribution with unknown density $f_0$. Let $\mathcal{N} = \{\mathcal{F}_g : g \in [\gamma]\}$ be some countable collection of polygonal density models up with component sizes up to $\gamma \in \mathbb{N}$. Let $\hat{f}_n^g \in \mathcal{F}_g$ be the ML estimator in the sense that $\sup_{f \in \mathcal{F}_g} L(t; \mathbf{X}_n) = L(\hat{f}_n^g; \mathbf{X}_n)$, where $L(f; \mathbf{X}_n) = \sum_{j=1}^n \log f(X_j)$ is the log-likelihood function indexed by $t$. There exists universal constants $\kappa$, $C_5$, and $C_6$, such that if we define the penalized ML criterion to be*

$$\text{crit}(g) = L\left(\hat{f}_n^g; \mathbf{X}_n\right) + \text{pen}(g), \tag{9}$$

*where*

$$\text{pen}(g) \geq \kappa \left[\left(\frac{C_5 g^{1/2}}{\left[n^{1/2} - C_4^{1/2}(g)\right]^3}\right)^{8/3} + \frac{g}{n}\right], \tag{10}$$

*then when the random $\hat{g}$ that minimizes crit over $\mathcal{N}$ exists, we have*

$$E\left[H^2\left(f_0, f_n^{\hat{g}}\right)\right] \leq C_6 \left[\inf_{g \in [\gamma]} \left[\inf_{f \in \mathcal{F}_g} K(f_0 \| f) + \text{pen}(g)\right] + \frac{\Sigma_\gamma}{n}\right],$$

*regardless of whether $t_0$ is in $\mathcal{N}$ or not.*

*Remark* 4. Theorem 3 provides a qualitative guarantee that as the sample size increases, we obtain smaller bounds on the expected Hellinger divergence (with respect to the KL divergence) when sample sizes increase, if we choose to select between polygonal distributions with up to $\gamma$ components via the minimization of the criterion (9). If $f_0$ is indeed in $\mathcal{N}$, then $\inf_{f \in \mathcal{F}_g} K(f_0 \| f)$ will go to zero as $n$ goes to infinity and thus the theorem also states that we have Hellinger consistency via the penalization criterion (9) as well. Finally, if we simply restrict $\mathcal{N} = \mathcal{F}_\gamma$ for some $\gamma \in \mathbb{N}$, we can use Theorem 3 to obtain an oracle-like result for the estimation of a singular class of models (cf. the remarks from Blanchard et al. (2008, Sec. 4.1)). Not that Theorem 3 is not technically an oracle result as it is not bounding the Hellinger divergence by itself, but rather the KL divergence instead.

*Remark* 5. Although Theorem 3 only provides a qualitative description on its own (i.e. because of the unknown constants in the penalty), it can be turned into a practical model selection technique via the use of the slope heuristic of Birge & Massart (2001) and Birge & Massart (2007). The slope heuristic can be applied to through the popular software package of Baudry et al. (2012). Implementations of the slope heuristic include the selection of $k$ in $k$-means (Fischer, 2011) and the number of components $g$ in a Gaussian mixture model (Baudry et al., 2012). In order to apply the method of (Baudry et al., 2012), one needs to first obtain a penalty that is known up to a constant multiple. We can for instance take $\kappa' = \kappa C_5^{8/3}$ and set the penalty to be

$$\text{pen}'(g) = \kappa' \left[ \left( \frac{g^{1/2}}{\left[ n^{1/2} - C_4^{1/2}(g) \right]^3} \right)^{8/3} + \frac{g}{n} \right].$$